\documentclass[]{interact}

\usepackage{epstopdf}
\usepackage{subfigure}

\usepackage{natbib}
\bibpunct[, ]{(}{)}{;}{a}{}{,}
\renewcommand\bibfont{\fontsize{10}{12}\selectfont}

\theoremstyle{plain}

\theoremstyle{definition}

\theoremstyle{remark}

\begin{document}


\title{Eddy saturation in a reduced two-level model of the atmosphere}

\author{
\name{Melanie Kobras\textsuperscript{a,b}\thanks{CONTACT Melanie Kobras. Email: m.kobras@pgr.reading.ac.uk}, Maarten H. P. Ambaum\textsuperscript{c} and Valerio Lucarini\textsuperscript{a,b}}
\affil{\textsuperscript{a}Department of Mathematics and Statistics, University of Reading, Reading, UK; \textsuperscript{b}Centre for the Mathematics of Planet Earth, University of Reading, Reading, UK; \textsuperscript{c}Department of Meteorology, University of Reading, Reading, UK}
}

\maketitle

\begin{abstract}
Eddy saturation describes the nonlinear mechanism in geophysical flows whereby, when average conditions are considered, direct forcing of the zonal flow increases the eddy kinetic energy, while the energy associated with the zonal flow does not increase. Here we present a minimal baroclinic model that exhibits complete eddy saturation. Starting from Phillips' classical quasi-geostrophic two-level model on the beta channel of the mid-latitudes, we derive a reduced order model comprising of six ordinary differential equations including parameterised eddies. This model features two physically realisable steady state solutions, one a purely zonal flow and one where, additionally, finite eddy motions are present. As the baroclinic forcing in the form of diabatic heating is increased, the zonal solution loses stability and the eddy solution becomes attracting. After this bifurcation, the zonal components of the solution are independent of the baroclinic forcing, and the excess of heat in the low latitudes is efficiently transported northwards by finite eddies, in the spirit of baroclinic adjustment.
\end{abstract}


\section{Introduction}
The equilibrium volume transport of the Antarctic Circumpolar Current (ACC) is known to be balanced by three contributions, namely the input of momentum at the ocean surface by the wind, the downward transport of this momentum by eddies and the bottom drag as the opposite force to the wind \citep{Munk1951Note, Nadeau2015TheRole}. \cite{Straub1993OntheTransport} was first to suggest that the ACC volume transport is independent of the wind stress at the ocean surface although the wind was assumed to be sufficiently strong. Since then more studies, using resolved turbulent ocean eddies, have confirmed this finding (e.g. \cite{Munday2013Eddy, Nadeau2015TheRole}). According to those studies, an increase in wind stress yields an increase of the vertical shear in the channel and the flow becomes baroclinically unstable, thus generating stronger eddies instead of a higher volume transport, a process called \textit{eddy saturation}.

\cite{Marshall2017Eddy}, inspired by a previous model of variability in atmospheric storm tracks \citep{Ambaum2014Anonlinear}, explained the physical principles of eddy saturation using a simple model with just three ingredients, including a zonal momentum budget, a closure relation between the eddy form stress and eddy energy, and an eddy energy budget. In their model, both the vertical shear and the volume transport of the ACC are predicted to be independent of the forcing by the surface wind and instead controlled by the requirement of a sufficiently unstable vertical shear to overcome the stabilizing role of the eddy energy dissipation. Moreover, the model explains the increase of eddy energy with wind stress and they conclude an analogy to the interaction between wave activity and baroclinicity in the original model of atmospheric storm tracks \citep{Ambaum2014Anonlinear}.

However, the process driving the variability of storm tracks in the mid-latitude atmosphere differs from the transport in the ocean in several ways. While mechanical stress dominates the oceanic processes described above, the baroclinic instability in the atmosphere is a result of the horizontal temperature gradient between the equator and poles, which is primarily due to the presence of a stronger radiative forcing at low rather then high latitudes. At all levels of the atmosphere, the meridional temperature gradient is proportional to the vertical shear of the mean flow by thermal wind balance and is the source of available potential energy for the eddies which are associated with the storm tracks. The corresponding poleward eddy heat fluxes in the storm tracks act to weaken the baroclinicity and therefore the mean flow (e.g. \cite{Pedlosky1979Geophysical, Holton2013Dynamic}). The latter process is often referred to as baroclinic adjustment \citep{Stone1978Baroclinic}. The Lorenz energy cycle provides a comprehensive view of energetics of the climate system that takes into account forcing, dissipation, and exchange of energy between available potential and kinetic form, and between energy pertaining to the mean flow and energy pertaining to eddy motions \citep{Lorenz1967Thenature,Peixoto1992Physics,Lucarini2014Mathematical}.

The heuristic model proposed by \cite{Ambaum2014Anonlinear} describes this baroclinic interaction between mean flow and eddy activity. In its steady state, the model predicts a two-way equilibration of storm tracks to extratropical thermal forcing and eddy friction: baroclinicity is independent of the thermal forcing but proportional to the eddy dissipation, whereas storm track activity is independent of eddy dissipation but proportional to thermal forcing of large-scale baroclinicity and is therefore reminiscent of the eddy saturation phenomenon in the ACC \citep{Novak2018Baroclinic}.

Within the Lorenz energy cycle framework, one can see the eddy saturation mechanism as that a forcing acting on the zonal fields results, on average, in an increase in the reservoir of eddy energy (in both kinetic and available potential form), whereas the reservoir of zonal energy stays largely unaffected. Additionally, one can view the eddy saturation mechanism as an extreme form of the baroclinic adjustment process; see also discussion in \cite{Lucarini2007Parametric}.

The simplest model that can incorporate baroclinic processes together with diabatic heating and surface friction is Phillips' two-level quasi-geostrophic model on the $\beta$-plane \citep{Phillips1956Thegeneral}. The present paper uses this model to derive a set of ordinary differential equations that are able to provide a minimal yet meaningful model of the above described interaction between mean flow and eddy activity. The novelty of this work is how the stability properties of the derived models are determined. In contrast to the usual approach of normal mode instability analysis (see e.g. \cite{Pedlosky1979Geophysical, Hoskins2014Fluid}), which focuses on studying the stability properties of the zonal flow, we use here methods of dynamical systems theory, which allow us to show the existence of a second attracting steady state instead of growing normal mode  baroclinic instabilities. Another novel aspect is that this second steady state exhibits the above described eddy saturation properties in a model with parameterised eddies.

In section \ref{sec:model_equations} we briefly review the two-level model and its reduction which follows work by \cite{Phillips1956Thegeneral} and by \cite{Thompson1987Large}. In section \ref{sec:non-dim_model} we introduce the non-dimensionalisation used to define our reduced model and determine the steady states. The stability and dependence on relevant parameters thereof is analysed and physically explained in section \ref{sec:stability} and compared to the normal mode baroclinic instability analysis of Phillips' two-level model in section \ref{sec:wave_cut-off}. In section \ref{sec:Q_factor} a quality factor is introduced to describe the oscillatory behaviour of the model and section \ref{sec:discussion} summarizes the results and discusses them in comparison to the current literature.

\section{The model equations}\label{sec:model_equations}

We use the two-level quasi-geostrophic model in pressure coordinates of \cite{Phillips1956Thegeneral}, consisting of two vorticity equations, coupled by a thermodynamic energy equation and including surface friction and a $\beta$-plane approximation, i.e. $f=f_0+\beta y.$ The equations are
\begin{subequations}
\begin{align}
    \left(\frac{\upartial}{\upartial t} + \bm{V}_1\cdot\nabla\right)(\beta y+\nabla^2\psi_1) - f_0\frac{\omega_2}{\delta p} &= A\nabla^2\nabla^2\psi_1,
    \label{eq:PV1} \\
    \left(\frac{\upartial}{\upartial t} + \bm{V}_3\cdot\nabla\right)(\beta y+\nabla^2\psi_3) + f_0\frac{\omega_2}{\delta p} &= A\nabla^2\nabla^2\psi_3 -\kappa\nabla^2\psi_4, \label{eq:PV2}\\
     \left(\frac{\upartial}{\upartial t}+\bm{V}\cdot\nabla\right)\frac{\psi_3-\psi_1}{\delta p} + \frac{R\overline{\Gamma}_2}{p_2f_0}\omega_2 &= -\frac{R}{p_2f_0c_{\rm p}}J_2, \label{eq:Thermo}
\end{align}
\end{subequations}
where $\psi$ is the streamfunction, ${\bm V}=(u,v)$ the horizontal velocity vector, $\omega={\rm d}p/{\rm d}t$ the vertical velocity and $\nabla$ is the horizontal gradient. The pressure is denoted by $p$, where $\delta p=500\,\text{hPa}$ is the pressure difference between the two vorticity levels as well as the pressure-thickness of each layer. Subscripts 1, 2, 3 and 4 denote the pressure levels 250, 500, 750, and 1000~hPa respectively. The fields are defined in the spatial domain $(x,y)\in[0,L]\times[0,W]$, where $L$ and $W$ are the length and width of the $\beta$-plane channel. We remark that, as customary, the $x$ coordinate is aligned with the longitude and the $y$ coordinate is aligned with the latitude. The fields above are defined for non-negative times $t\geq 0$ starting from smooth initial conditions. Clearly, we have that $A(0,y,t)=A(L,y,t)$ $\forall y\in[0,W]$ and $\forall t\geq0$ for all fields $A$. 

The parameters $A$ and 
$\kappa$ describe the eddy diffusion and the surface friction diffusion. 
Additionally, $J$ is the diabatic heating and $\overline{\Gamma}=RT/pc_{\rm p}-\upartial T/\upartial p$ the basic state static stability, where $T$ is the temperature, $c_{\rm p}$ the isobaric specific heat capacity of dry air and $R$ the specific gas constant for dry air. 

The diabatic heating $J$ is given by the sum of 
radiative and diffusive contributions. The radiative contribution represents the net local radiative energy gains and losses and is chosen to vary linearly in $y,$
\begin{equation}
J_{\text{rad}}= 2H\left(1-\frac{2y}{W}\right),\label{eq:Jrad}
\end{equation}
where $H$ is the mean rate of heating per unit mass for $y\in[0,W/2]$ (or cooling for $y\in[W/2,W]$). The diffusive contribution represents the effect of lateral eddy diffusion of temperature at level 2:
\begin{equation}
    J_{\text{diff}}= c_{\rm p}A\nabla^2T_2 = \frac{2f_0c_{\rm p}p_2}{\delta p R}A\nabla^2\psi_{\rm T}.\label{eq:Jdiff}
\end{equation}
We remark that the parameters $A$ and $\kappa$ control the dissipation processes that remove energy from the system. The parameter $H$ controls the input of energy in the form of an increase in temperature difference between low and high latitudes, hence fuelling the Lorenz energy cycle of the system by production of zonal mean available potential energy. The competition between dissipation and forcing determines the degree of instability and turbulence of the system, which, after transients, reaches a statistically steady state, the \textsl{model climate}.

Next, we define barotropic and baroclinic potential vorticities $q_{\rm m}=\nabla^2\psi_{\rm m}+\beta y$ and $q_{\rm T}=(\nabla^2-\lambda^2_{\rm R})\psi_{\rm T}$, where $\lambda^2_{\rm R}=2p_2f^2_0/(\delta p^2\overline{\Gamma}_2R)$ is the inverse square of the Rossby radius of deformation and the barotropic and baroclinic streamfunctions are defined as $\psi_{\rm m}=(\psi_1+\psi_3)/2$ and $\psi_{\rm T}=(\psi_1-\psi_3)/2$. 

Since we want to obtain evolution equations for the longitudinally averaged or mean flow (denoted by overbars) and longitudinally varying disturbances or eddies (denoted by primes) separately, we introduce the zonal mean $\overline{A}(y,t)=1/L\int_0^L\mathrm{d}xA(x,y,t)$ and define the deviation from such mean as $A'(x,y,t)=A(x,y,t)-\overline{A}(y,t)$ for any field $A$. 

Now the evolution equations for the mean zonal wind and mean thermal wind or shear are obtained by taking half the sum and the difference of equations (\ref{eq:PV1}) and (\ref{eq:PV2}) where we introduced the geostrophic assumption of horizontally non-divergent flow, yielding the geostrophic relation $u=-\upartial\psi/\upartial y$ and $v=\upartial\psi/\upartial x$ for the velocities in all terms but the one containing the Coriolis parameter. Substituting equation (\ref{eq:Thermo}), taking the derivative with respect to $y$ and applying the definitions and assumptions above, we arrive at
\begin{subequations}
\begin{align}
    &\frac{\upartial \overline{u}_{\rm m}}{\upartial t} -\frac{1}{\lambda_y^2}  \frac{\upartial^3}{\upartial y^3}\overline{v'_{\rm m} u'_{\rm m}}   -\frac{1}{\lambda_y^2}\frac{\upartial^3}{\upartial y^3}\overline{v'_{\rm T} u'_{\rm T}} = -\lambda_y^2 A\overline{u}_{\rm m} - \frac{\kappa}{2}(\overline{u}_{\rm m}-2\overline{u}_{\rm T}), \label{eq:u_m_eddy}\\
    &\frac{\upartial \overline{u}_{\rm T}}{\upartial t} -  \frac{1}{\lambda_y^2+\lambda_{\rm R}^2} \frac{\upartial^3}{\upartial y^3}\overline{v'_{\rm T} u'_{\rm m}} - \frac{1}{\lambda_y^2+\lambda_{\rm R}^2} \frac{\upartial^3}{\upartial y^3}\overline{v'_{\rm m} u'_{\rm T}} -\frac{\lambda^2_{\rm R}}{\lambda^2_y+\lambda_{\rm R}^2}\frac{\upartial^2}{\upartial y^2}\overline{v'_{\rm m}\psi'_{\rm T}} \label{eq:u_T_eddy}\\
    &\qquad \qquad \qquad \qquad\qquad \qquad = -\lambda^2_y A\overline{u}_{\rm T} + \frac{\lambda_y^2}{\lambda_y^2+\lambda_{\rm R}^2}\frac{\kappa}{2}(\overline{u}_{\rm m}-2\overline{u}_{\rm T}) + \frac{\lambda^2_{\rm R}}{\lambda_y^2+\lambda_{\rm R}^2}\frac{2\delta pRH}{f_0p_2c_{\rm p}W}, \nonumber
\end{align}
\end{subequations}
where $\overline{u}_{\rm m}$ is the mean zonal wind and $\overline{u}_{\rm T}$ the mean thermal wind or mean zonal shear. Here we additionally assumed a specific meridional shape of the mean wind, namely $\upartial^2\overline{u}/\upartial y^2=-\lambda_y^2\overline{u}$ where $\lambda_y^2$ is an unspecified wavenumber.

Following \cite{Thompson1987Large}, we next restrict the eddy component of the barotropic and baroclinic streamfunction to be of the form
\begin{equation}
    \psi'_{\rm m,T}(x,y,t)=A_{\rm m,T}(t)\sin kx\sin ly+B_{\rm m,T}(t)\cos kx\sin ly, \label{eq:Eddy_shape}
\end{equation}
where the wavenumber $k$ remains unspecified and $l=\pi/W$. We note that $\nabla^2\psi'_{\rm m,T}=-\lambda^2_{\nabla}\psi'_{\rm m,T}$ with $\lambda^2_{\nabla}=k^2+l^2=k^2+\pi^2/W^2$. Using this assumption, all eddy product terms in equations (\ref{eq:u_m_eddy}) and (\ref{eq:u_T_eddy}) besides the net poleward eddy temperature flux $v'_{\rm m}\psi'_{\rm T}$ cancel. Using the explicit expression (\ref{eq:Eddy_shape}) for $v'_{\rm m}$ and $\psi'_{\rm T}$ we obtain
\begin{subequations}
\begin{align}
    \overline{v'_{\rm m}\psi'_{\rm T}} &= \frac{k}{2}(A_{\rm m}B_{\rm T}-A_{\rm T}B_{\rm m})\sin^2\frac{\pi y}{W}, \label{eq:Eddy_shape_1}\\
    \frac{\upartial^2}{\upartial y^2}\overline{v'_{\rm m}\psi'_{\rm T}} &= \frac{\pi^2k}{W^2}(A_{\rm m}B_{\rm T}-A_{\rm T}B_{\rm m})\left(\cos^2\frac{\pi y}{W}-\sin^2\frac{\pi y}{W}\right), \label{eq:Eddy_shape_2}
\end{align}
\end{subequations}
so at $y=W/2$,
\begin{equation}
    \frac{\upartial^2}{\upartial y^2}\overline{v'_{\rm m}\psi'_{\rm T}} = -\frac{2\pi^2}{W^2}\overline{v'_{\rm m}\psi'_{\rm T}}.\label{eq:Eddy_shape_3}
\end{equation}
We see that the net poleward heat transport attains its maximum at $y=W/2$ and vanishes at $y=0$ and $y=W$. Indeed, this qualitatively corresponds to what is observed in more complex versions of the same model \citep{Lucarini2007Parametric} and to what is observed in the actual climate \citep{Peixoto1992Physics}. Following again \cite{Thompson1987Large}, we assume that the mean thermal wind $\overline{u}_{\rm T}$ has the same meridional structure, and inspecting equation (\ref{eq:u_m_eddy}) we see that the same must be true for the mean zonal wind $\overline{u}_{\rm m}$. Additionally, from the assumption about the zonal mean wind, it implies that $\lambda^2_y=2\pi^2/W^2$. Finally, equations (\ref{eq:u_m_eddy}) and (\ref{eq:u_T_eddy}) can be substantially simplified to
\begin{subequations}
\begin{align}
    \frac{\upartial \overline{u}_{\rm m}}{\upartial t} &= -\lambda_y^2 A\overline{u}_{\rm m} - \frac{\kappa}{2}(\overline{u}_{\rm m}-2\overline{u}_{\rm T}), \label{eq:u_m_final}\\
    \frac{\upartial \overline{u}_{\rm T}}{\upartial t} &= - \frac{\lambda_y^2\lambda_{\rm R}^2}{\lambda_y^2+\lambda_{\rm R}^2} \overline{v'_{\rm m} \psi'_{\rm T}} -\lambda^2_y A\overline{u}_{\rm T} + \frac{\lambda_y^2}{\lambda_y^2+\lambda_{\rm R}^2}\frac{\kappa}{2}(\overline{u}_{\rm m}-2\overline{u}_{\rm T}) + \frac{\lambda^2_{\rm R}}{\lambda_y^2+\lambda_{\rm R}^2}\frac{2\delta pRH}{f_0p_2c_{\rm p}W}, \label{eq:u_T_final}
\end{align}
\end{subequations}
which is now a closed system by an evolution equation for the net poleward temperature flux.

This equation and evolution equations for three other occurring eddy correlation terms, namely the mean meridional kinetic energy $\overline{v'^2_{\rm m}}$, the temperature variance $\overline{v'^2_{\rm T}}$ and the cross-correlation between temperature and geopotential $\overline{v'_{\rm m}v'_{\rm T}}$, are derived following to some extend the technique by \cite{Thompson1987Large}, with the difference that our model includes a surface friction term. The poleward temperature flux equations is
\begin{align}
    \frac{\upartial}{\upartial t}\overline{v'_{\rm m}\psi'_{\rm T}} &=  \frac{\lambda^2_{\rm R}+\lambda^2_y-\lambda^2_{\nabla}}{\lambda^2_{\nabla}+\lambda^2_{\rm R}}\overline{u}_{\rm T}\overline{v'^2_{\rm m}} - \frac{ \lambda^2_{\rm R}}{\lambda^2_{\nabla}(\lambda^2_{\nabla}+\lambda^2_{\rm R})}\left(\beta+\lambda^2_y\overline{u}_{\rm m}\right)\overline{v'_{\rm m}v'_{\rm T}} + \frac{\lambda^2_{\nabla}-\lambda^2_y}{\lambda^2_{\nabla}}\overline{u}_{\rm T}\overline{v'^2_{\rm T}} \nonumber\\ 
    &\qquad - \left(2A\lambda^2_{\nabla}+\frac{\kappa(3\lambda^2_{\nabla}+\lambda^2_{\rm R})}{2(\lambda^2_{\nabla}+\lambda^2_{\rm R})}\right)\overline{v'_{\rm m}\psi'_{\rm T}}.\label{eq:v_m_psi_T_final}
\end{align}
The evolution equations for the three other eddy correlation terms can be found in appendix \ref{sec:Additional_equations}.

\section{The non-dimensional model}\label{sec:non-dim_model}
Up to this point the meridional wavenumber $k$ of the eddy component of the streamfunctions remained unspecified. In principle, one has that $k=2p\pi/L$, where $p$ is a non-vanishing natural number. Now, we let $k$ be a multiple of the longitudinal wavenumber $l$ and we define $k=jl=j\pi/W$ where $j$ is a parameter describing the aspect ratio of the eddies, that is -- for simplicity -- assumed to be a non-negative number. Indeed, $j$ can only assume discrete values, but since $L\gg W$, they are closely spaced and so the discrete nature of $j$ is neglected in what follows. Hence, $j$ larger than 1 means that the eddies are elongated in the meridional direction. If $j$ is smaller than 1 but positive, the eddies are elongated in the longitudinal direction.

To simplify the notation, we introduce the dimensionless variables
\begin{equation*}
    M=\frac{\lambda^2_{\rm R}}{2\beta}\overline{u}_{\rm m}, \ S=\frac{\lambda^2_{\rm R}}{2\beta}\overline{u}_{\rm T}, \ T=\frac{\lambda^5_{\rm R}}{2^3\beta^2}\overline{v'_{\rm m}\psi'_{\rm T}}, \
    K=\frac{\lambda^4_{\rm R}}{2^2\beta^2}\overline{v'^2_{\rm m}}, \ V=\frac{\lambda^4_{\rm R}}{2^2\beta^2}\overline{v'^2_{\rm T}}, \ X=\frac{\lambda^4_{\rm R}}{2^2\beta^2}\overline{v'_{\rm m}v'_{\rm T}}
\end{equation*}
and define the time variable $\tau=t\beta/\lambda_{\rm R}$, so that equations (\ref{eq:u_m_final}), (\ref{eq:u_T_final}), (\ref{eq:v_m_psi_T_final}) and (\ref{eq:v^2_m})-(\ref{eq:v_m_v_T}) take the form
\begin{subequations}
\begin{align}
     \frac{{\rm d}M}{{\rm d}\tau}&= -aM + \alpha S, \label{eq:dMdt}\\
     \frac{{\rm d}S}{{\rm d}\tau}&= \frac{\gamma\alpha}{2}M - bS - 4\gamma T +\eta H, \label{eq:dSdt}\\
     \frac{{\rm d}T}{{\rm d}\tau}&= \mu SK - \frac{\delta\lambda^2_{\rm R}}{2\lambda_y^2}X - \delta MX - cT  + \frac{j^2-1}{j^2+1}SV, \label{eq:dTdt}\\
     \frac{{\rm d}K}{{\rm d}\tau}&= \frac{2^2j^2(j^2-1)}{j^2+1}\frac{\lambda^2_y}{\lambda^2_{\rm R}}ST-dK +2\alpha X, \label{eq:dKdt}\\
     \frac{{\rm d}V}{{\rm d}\tau}&= \frac{2^2j^2\lambda^2_y\mu}{\lambda^2_{\rm R}} ST - eV + \alpha\zeta X, \label{eq:dVdt}\\
     \frac{{\rm d}X}{{\rm d}\tau}&= j^2\delta\left(1+\frac{2\lambda^2_y}{\lambda^2_{\rm R}}M\right)T-cX +\alpha V +\frac{\alpha\zeta}{2}K\label{eq:dXdt},
\end{align}
\end{subequations}
where the definitions of the dimensionless constants $\alpha$, $\gamma$, $\delta$, $\epsilon$, $\nu$ and $\eta$ can be found in appendix \ref{sec:dim-less constants}.

For a non-zero heating rate $H$ this system exhibits three steady states. The first one is a zonal steady state where all eddy components are zero:
\begin{equation}
    P^0 = (M^0,S^0,0,0,0,0) = \frac{2}{2ab-\gamma \alpha^2}\eta H(\alpha,a,0,0,0,0). \label{eq:zonal_state}
\end{equation}
Here, $M^0$ and $S^0$ are always positive because the denominator of the pre-factor and all parameter values including the heating rate are positive within the considered parameter space (see appendix \ref{sec:Parameter_space}). The second steady state, denoted by $P^*=(M^*, S^*, T^*, K^*, V^*, X^*)$, is given by
\begin{align}
    M^* &= \frac{\alpha}{a}S^*,& 
    S^* &= \frac{j^2a}{A}\left(\left(B^2+4CD\right)^{1/2}+B\right),&
    T^* &= -\frac{2ab-\gamma\alpha^2}{8a\gamma}S^* + \frac{\eta}{4\gamma}H,& \label{eq:eddy_state}
\end{align}
where $K^*$, $V^*$, $X^*$ and the new parameters $B$, $C$ and $D$ can be found in appendix \ref{sec:eddy_steady_state}. Note that $B$, $C$ and $D$ are independent of the heating $H$ and therefore the steady state mean zonal wind and shear are independent of the heating.

The third steady state of the system can be excluded as unphysical since the values of $K$ and $V$ are negative in the given parameter space despite being non-dimensional variables for $\overline{v_{\rm m}'^2}$ and $\overline{v_{\rm T}'^2}$, respectively, which being squared real quantities must always be non-negative. Therefore the system exhibits two physical solutions, in the following referred to as the zonal ($P^0$) and the eddy steady state ($P^*$).

\section{Physical mechanisms of zonal and eddy saturated steady state}\label{sec:stability}

The linear stability of these two solutions is determined by numerically calculating the eigenvalues of the Jacobian matrix of system (\ref{eq:dMdt})-(\ref{eq:dXdt}) at each steady state. Therefore, the parameters in table \ref{tab:Fixed_parameters} were kept fixed, whereas the surface friction diffusion $\kappa$, the eddy diffusion $A$, the mean rate of heating $H$ and the parameter $j$ describing the eddy aspect ratio were varied within the ranges given in table \ref{tab:Varying_parameters}.

In the following the notion of an attracting (stable) and repelling (unstable) steady state is used to describe the linear stability of the two steady states. Hence, if the eddy steady state is attracting, the model converges to the steady state with finite eddy contributions. This state then breaks the zonal symmetry of the zonal state: the phase of the eddies is not determined but the eddy correlation statistics are fixed in time. On the other hand, an attracting zonal steady state describes a state of the atmosphere where eddy activity decays until the flow is purely zonal, and once in that state remains zonal.

\begin{figure}
\begin{center}
\subfigure{
\resizebox*{7.1cm}{!}%
{\includegraphics{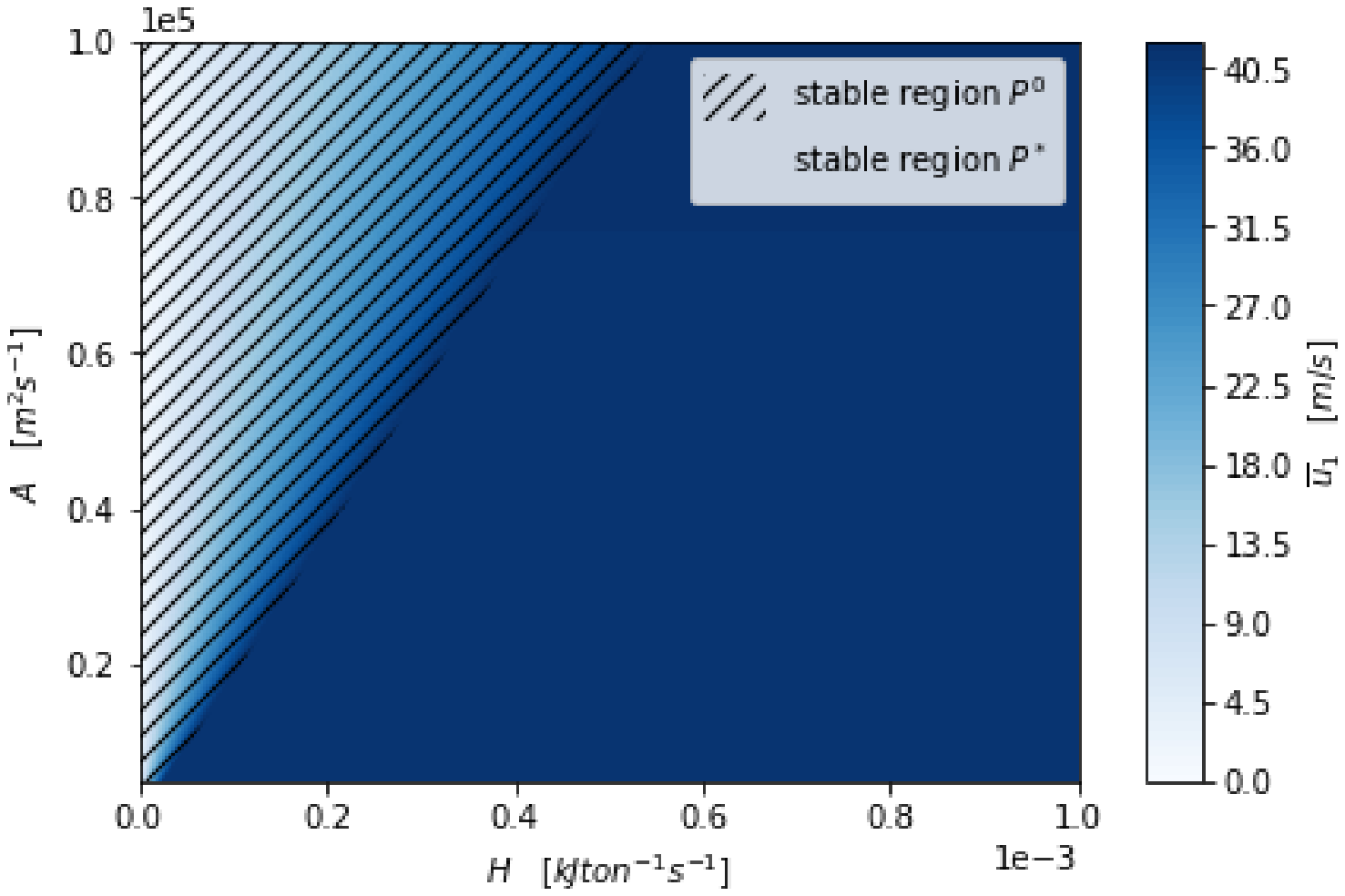}}}%
\subfigure{
\resizebox*{7.1cm}{!}%
{\includegraphics{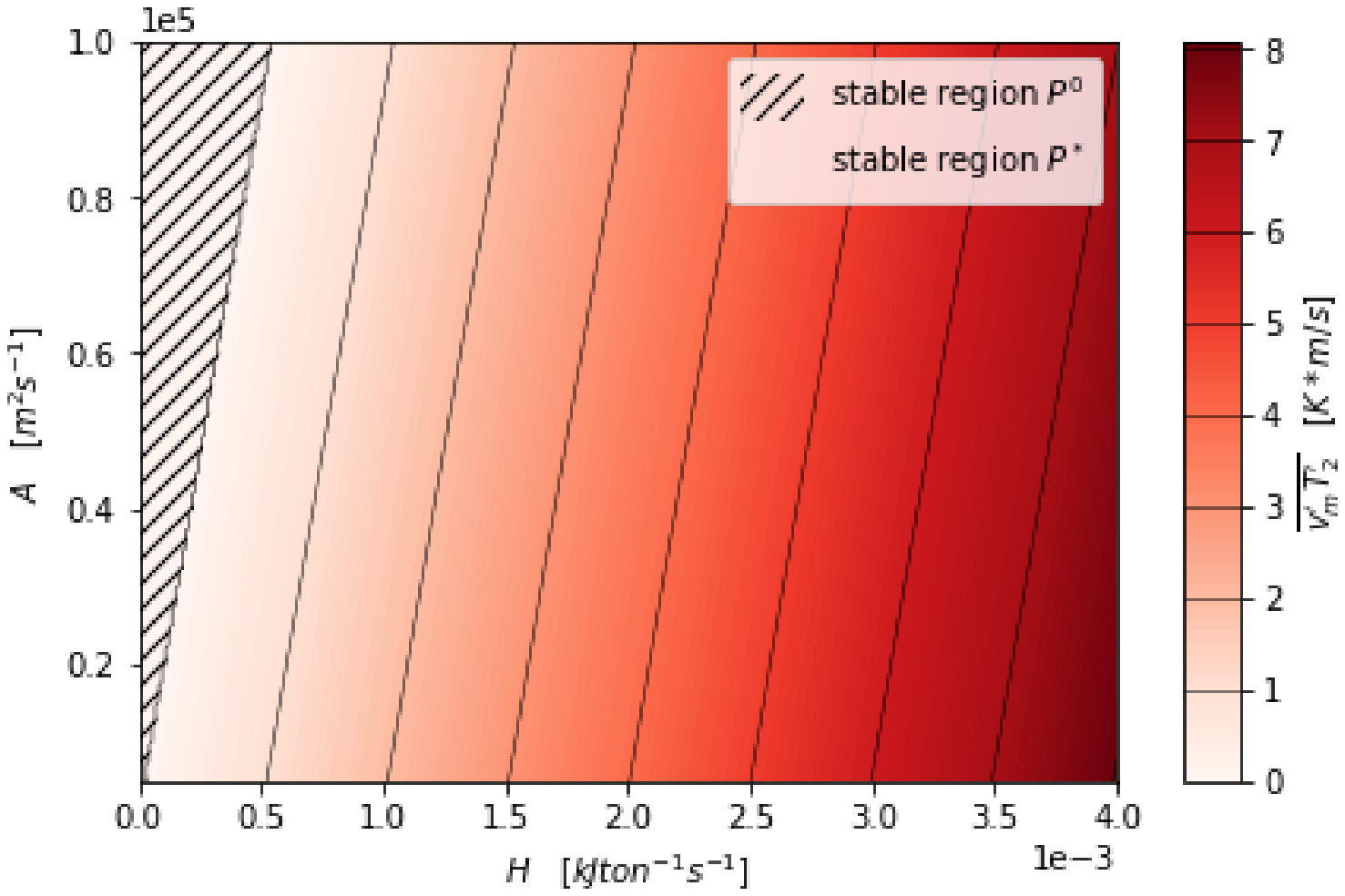}}}%
\caption{\label{fig:H_Avisc}
Contours of the mean zonal wind at the top level $\overline{u}_1$ (left) and the net poleward eddy temperature flux (right) for the respective attracting steady state; the dashed region is the parameter space where the zonal steady state $P^0$ is attracting, the non-dashed region where the eddy steady state $P^*$ is attracting; note that the range of $H$ in the right figure is four times the range in the left; $\kappa=4\times10^{-6}\text{ s}^{-1}$, $j=2.5$.}
\end{center}
\end{figure}

The considered parameter space is divided into two opposite stability regions where either the zonal or the eddy steady state is attracting and the respective other state is repelling. Keeping $\kappa$, $A$ and $j$ fixed but increasing the heating rate $H,$ a transcritical bifurcation occurs (see for example \cite{Guckenheimer1983Nonlinear}) and the system switches stability from the zonal to the eddy steady state. From equation (\ref{eq:zonal_state}) it is clear that the mean zonal wind and shear of $P^0$ grow linearly with the heating $H$. However, this proportionality is lost for the eddy steady state, where $M^*$ and $S^*$ are independent of the heating rate. This can be seen directly from equation (\ref{eq:eddy_state}) and from the left hand side of figure \ref{fig:H_Avisc}, where contours of the mean zonal wind at the top level for the respective attracting steady state are shown. Such an insensitivity of the mean zonal wind to the forcing by the heating rate closely resembles the  eddy saturation mechanism discussed in the introduction. 

In contrast to that, the (non-dimensional) net poleward heat transport $T^*$ grows linearly with $H$ (see again equation (\ref{eq:eddy_state}) and figure \ref{fig:H_Avisc} on the right) and is non-negative in its attracting parameter region only. For values of $H$ where the zonal steady state is attracting, $T^*$ is negative and therefore the net eddy heat transport is equatorward for the eddy state. This would imply that the system transports heat from cold towards warm regions, which suggests a condition that is thermodynamically not realisable. This further clarifies that the eddy steady state is physically irrelevant in its repelling region. Furthermore, it is clear from equation (\ref{eq:eddy_state}) that the steady state value of the net poleward heat transport only depends on the shear and incoming heat but not on the three other eddy correlation variables $K^*$, $V^*$ and $X^*$. Hence, the dynamics of the system are resembled by the mean zonal shear (proportional to the mean zonal wind) and the net poleward heat transport, and $K^*$, $V^*$ and $X^*$ do not need to be considered separately.

The above described stability switch is also dependent on the eddy diffusion parameter $A$. This is illustrated in figure \ref{fig:H_Avisc}, where a higher eddy diffusion yields a larger heating rate at which the zonal steady state loses stability. The relation between eddy diffusion and heating rate in $P^0$ is determined by equation (\ref{eq:zonal_state}). Replacing $S^0$ by the mean poleward temperature gradient at the middle level $\overline{T}_2$ yields
\begin{equation}
    G(A,\kappa)\left(-\frac{\upartial\overline{T}_2}{\upartial y}\right) = \frac{4\lambda^2_{\rm R}}{c_{\rm p}W\left(\lambda^2_y+\lambda^2_{\rm R}\right)}H \quad \text{with} \quad G(A,\kappa) = \lambda^2_y A\frac{2\lambda^2_y A +(2\gamma+1)\kappa}{2\lambda^2_y A +\kappa}. \label{eq:function_G}
\end{equation}
Since in the considered parameter space $\kappa$ is much smaller than one, the function $G$ is approximately linear in $A$. In physical terms, for a fixed equator to pole mean temperature gradient, a linear increase of the heating rate $H$ can be balanced by a linearly increased eddy diffusion. In this sense, the stabilization of the zonal state against heating is always realized by an eddy heat transport, be it parameterised, as an eddy diffusion, or explicit, as in eddy heat transport.

\begin{figure}
\begin{center}
\subfigure{
\resizebox*{7.1cm}{!}%
{\includegraphics{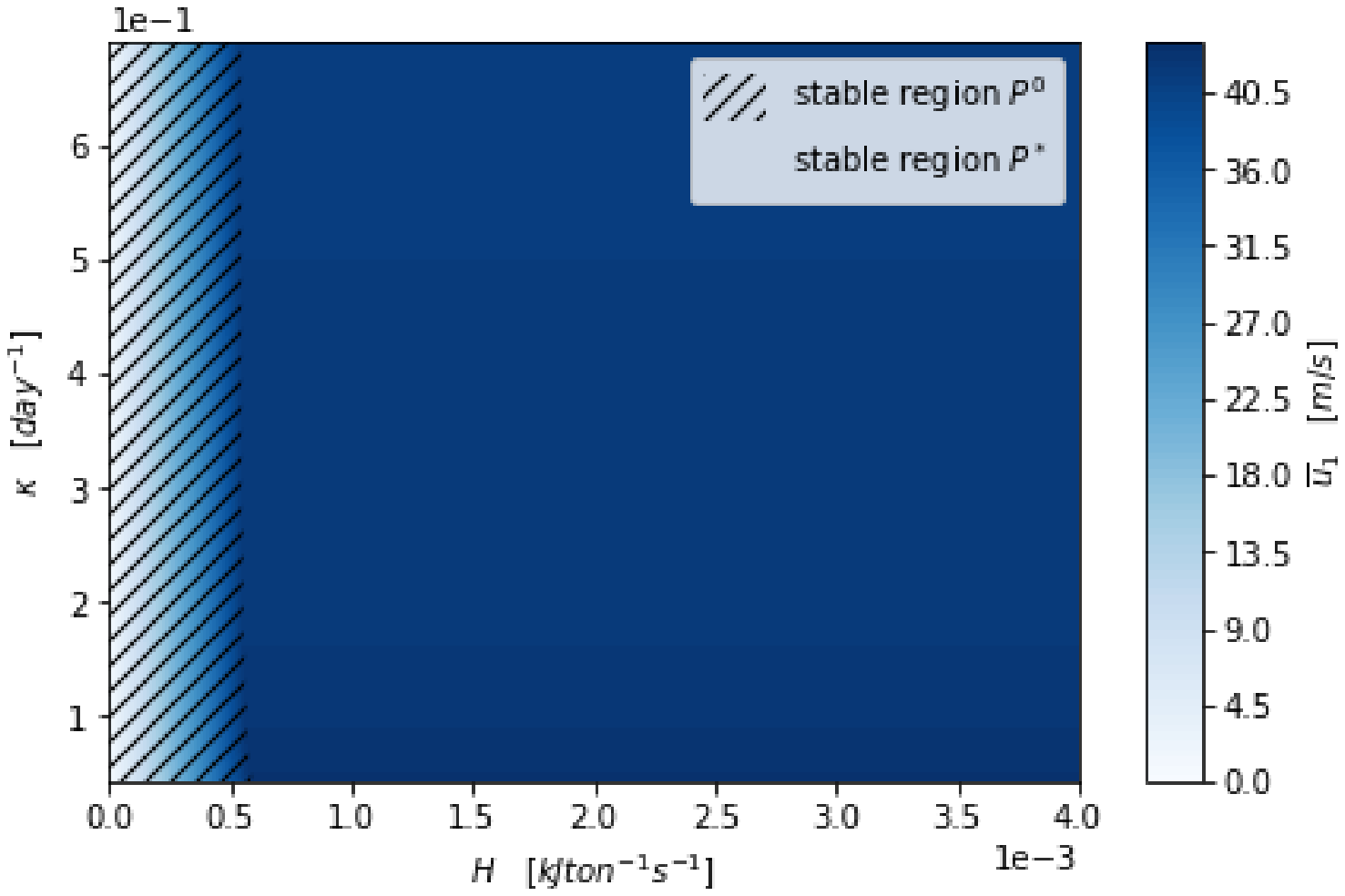}}}%
\subfigure{
\resizebox*{7.1cm}{!}%
{\includegraphics{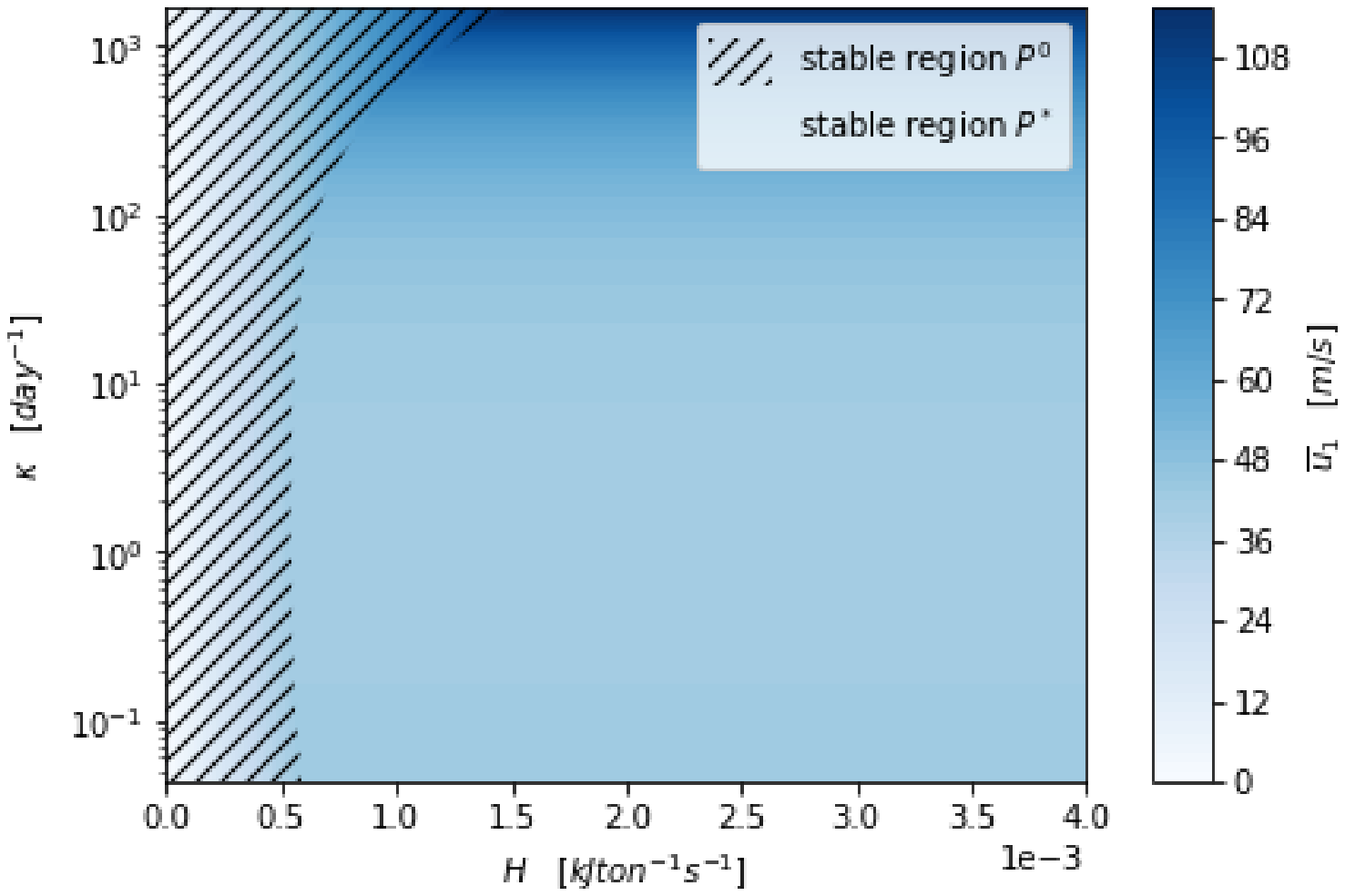}}}%
\caption{\label{fig:H_kappa}
Contours of the mean zonal wind at the top level $\overline{u}_1$ for realistic (left) and extremely high (right) values of the surface friction parameter $\kappa$; the dashed region is the parameter space where the zonal steady state $P^0$ is attracting, the non-dashed region where the eddy steady state $P^*$ is attracting; note that in the left figure the scale for the surface friction is logarithmic; $A=10^{5}\text{ m}^{2}\text{ s}^{-1}$, $j=2.5$.}
\end{center}
\end{figure}

It is also clear from (\ref{eq:function_G}) that the surface friction $\kappa$ is not able to balance the heating and the stability analysis shows that a change in $\kappa$ does not influence the critical value of heating for realistic values of $\kappa$ (see figure \ref{fig:H_kappa} on the left).

These opposed roles of the zonal and eddy steady state describe two physically related mechanisms of the atmosphere to compensate incoming heat. In the attracting region of the zonal steady state the relatively low heating can be compensated by the eddy diffusion. However, this balance is no longer achievable for increasing $H$ and decreasing $A$ so that in the eddy steady state the eddies have to reach a finite size to perform poleward transport of heat and 
compensate the atmospheric radiative energy imbalance. In the actual atmosphere -- or in more complex models -- this corresponds to the case where the flow is baroclinically unstable, and baroclinic cyclones grow and decay as manifestation of an active Lorenz energy cycle and, at the same time, transport heat poleward. Although it is a similar mechanism, the compensation of the heating does not depend on the friction diffusion $\kappa$. This is explained later in this section.

Another parameter that influences the stability regions of the steady states is the aspect ratio $j\in {\mathbb R}^+$ of the eddy shape. For the previous analysis $j=2.5$ was chosen, which corresponds to eddies elongated by a factor of $2.5$ in the meridional direction. For this aspect ratio, and even more meridionally elongated eddies, a stability switch always occurs and the larger $j$, the smaller is the heating rate $H$ at this switching point. In contrast to that, for $j$ smaller than one, corresponding to longitudinally elongated eddies, there is no exchange of stability and the zonal steady state is attracting for the whole parameter space under consideration.

\begin{figure}
\begin{center}
\subfigure{
\resizebox*{7cm}{!}%
{\includegraphics{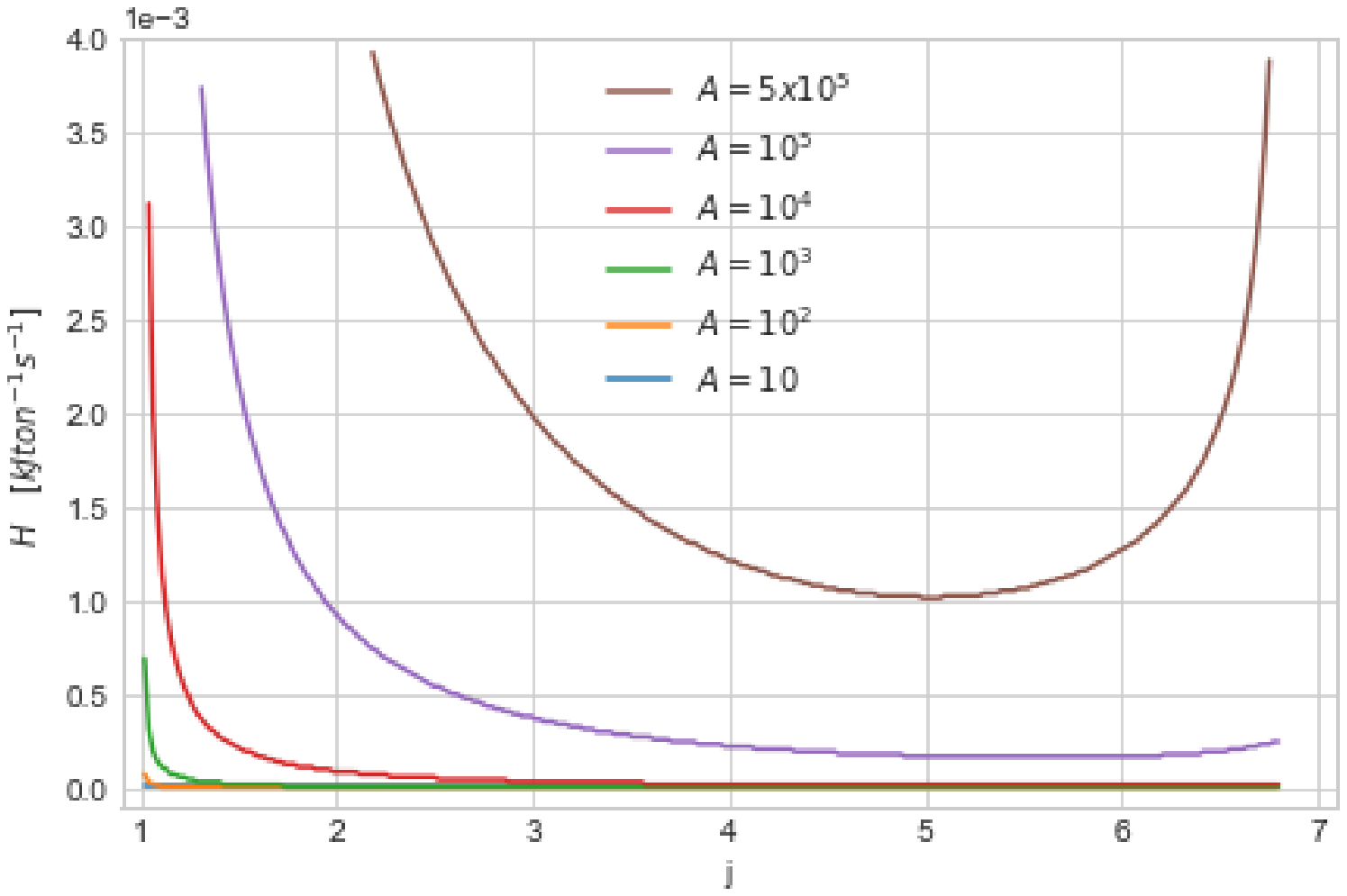}}}%
\subfigure{
\resizebox*{7cm}{!}%
{\includegraphics{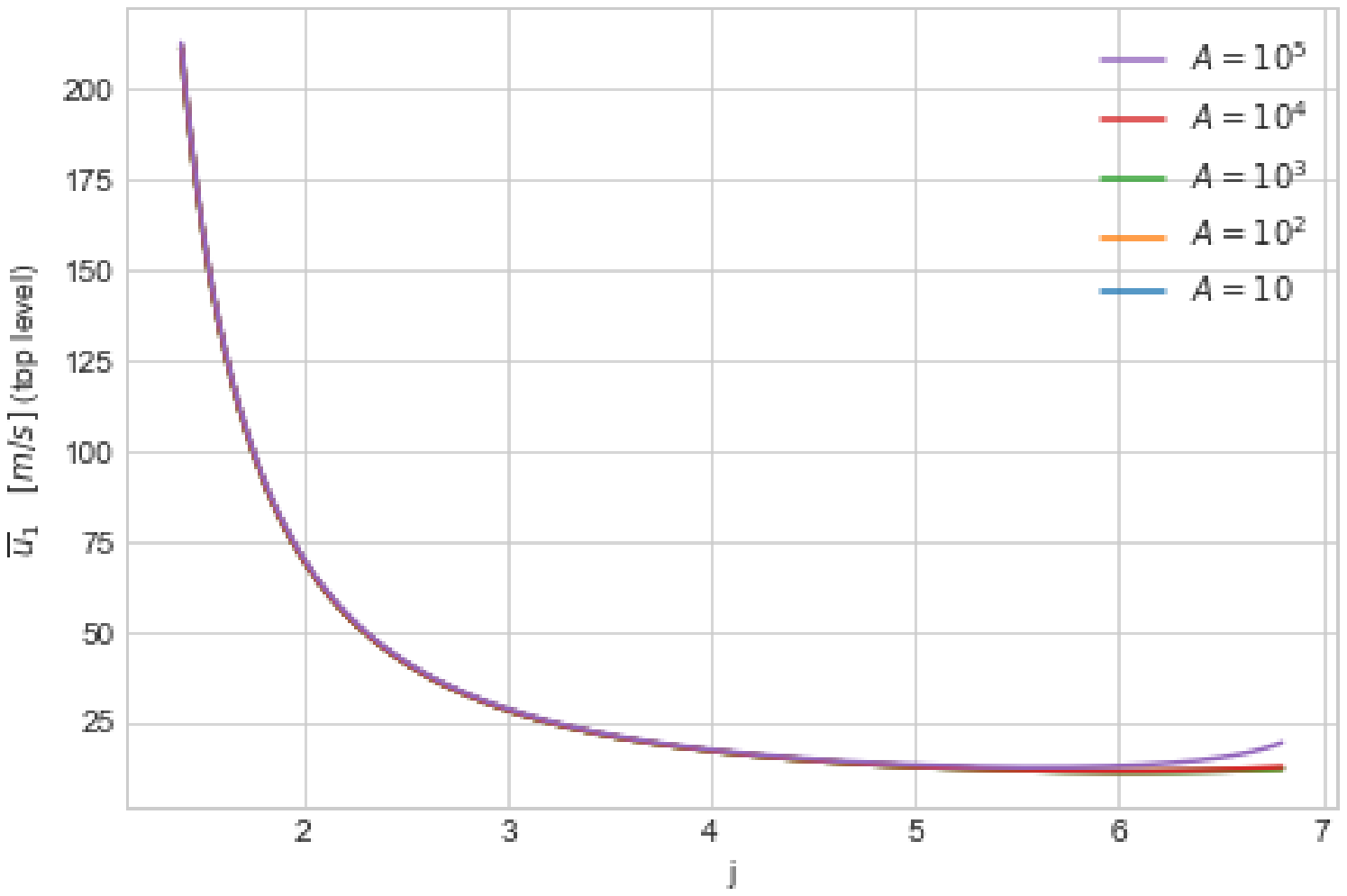}}}%
\caption{\label{fig:j_H_u1_Avisc}
The value of the heating rate $H$ at which the stability switch occurs (left) and the mean zonal wind at the top level $\overline{u}_1$ (right) as a function of the eddy shape aspect ratio $j$, shown for several values of the eddy diffusion $A$ in $\text{m}^{2}\text{ s}^{-1}$; $H=3.5\times10^{-3}\text{KJ ton}^{-1}\text{ s}^{-1}$ (right); $\kappa=4\times10^{-6}\text{ s}^{-1}$ (both).}
\end{center}
\end{figure}

For an aspect ratio above one, the value of $H$ at which the zonal state loses stability is additionally dependent on the eddy diffusion parameter. The left hand side of figure \ref{fig:j_H_u1_Avisc} shows the values of $H$ at which the switch occurs as a function of $j$ for several orders of magnitude of $A$. For a vanishing small eddy diffusion (e.g. $A$ on the order of $10\text{ m}^{2}\text{ s}^{-1}$) the eddy steady state is attracting for the whole range of $H$ and an aspect ratio larger than one. Increasing $A$ yields an exchange of stability for higher $H$, which was already seen in figure \ref{fig:H_Avisc}, and additionally an increasing value of $j$ is required for the eddy steady state to become attracting.

Hence, for meridionally more elongated eddies the eddy steady state becomes attracting at a lower heating rate. It is indeed well known that, in the full quasi-geostrophic flow, baroclinic instability is facilitated for these geometrical conditions \citep{Pedlosky1979Geophysical}. Only for a very large aspect ratio $j$ above five and for a large eddy diffusion the critical value of $H$ increases again. As before, besides this behaviour at very high aspect ratios, a change in the surface friction does not change this critical heating rate.

The right hand side of figure \ref{fig:j_H_u1_Avisc} shows the mean zonal wind at the top level of the eddy steady state as a function of the aspect ratio. Eddies close to a circular shape, i.e. $j$ close to one, yield unrealistically high wind speeds just above $215\text{ m }\text{s}^{-1}$. For more meridionally elongated eddies the wind speed decreases for an aspect ratio of up to 5 and thereafter increases again for very high eddy diffusion values. For $j\geq 2.5$ one gets reasonably realistic values for the wind speed. However, before this upward slope, the wind speed neither depends on the eddy diffusion nor on the surface friction diffusion (the latter not shown).

For the net poleward heat transport the opposite holds: the longer the eddies in the meridional direction, the higher the heat transport until it drops again for very large $j$. Besides for these very high aspect ratios, the surface friction diffusion $\kappa$ has again no impact on the values, whereas a very high eddy diffusion $A$ yields a decreased heat transport for the whole range of $j$.

As briefly mentioned earlier, the surface friction $\kappa$ could perhaps be thought to act similarly to the eddy diffusion $A$, taking out energy from the system and thus compensating for the heating $H$. However, the heating forces a vertical zonal shear by thermal wind balance, but would not force a zonal wind directly at the surface. The model can develop very low surface winds but increase its shear by increasing upper level zonal winds. In this sense, the surface friction does not directly compensate for the incoming heating $H$.

However, insofar that the heating induces surface winds, the surface friction would represent a sink for those surface winds. In that case, one would expect that a higher surface friction is able to dissipate the incoming energy and therefore moves the stability switch to higher values of the heating. This frictional adjustment can not be seen for realistic values of $\kappa$ (figure \ref{fig:H_kappa} on the left) but is observable for extremely high surface friction values (figure \ref{fig:H_kappa} on the right).



Additionally, and in contrast to the eddy diffusion $A$, an extremely high surface friction increases the mean zonal wind to unrealistically high values. This behaviour of the mean flow was also observed by \cite{Novak2018Baroclinic} and \cite{Marshall2017Eddy} for the Antarctic Circumpolar Current.

\section{Short and long wavelength cut-off}\label{sec:wave_cut-off}

For sufficiently large eddy diffusion $A$, the left hand side of figure \ref{fig:j_H_u1_Avisc} shows a short (large $j)$ and long wavelength (small $j$) cut-off of the instability of the zonal solution. In other words, our model only exhibits finite-size eddies within a certain range of wavelengths and the zonal flow stabilizes for very short and long waves.

A similar result follows from the classical stability analysis of the Phillips model presented in, for example, \cite{Holton2013Dynamic}. This is done for a simpler version of the two-level model, without surface friction, eddy diffusion or external forcing due to heating, and is obtained by performing a linear perturbation analysis. Assuming wave-like solutions for the barotropic and baroclinic development of the streamfunction, they obtain a dispersion relation for the phase speed of the waves and from this derive conditions for instability of the zonal flow and the growth of perturbations. The zonal flow is stable for very short waves because they are inefficient in converting available into potential energy, and for very long waves, because they are stabilized by the differential rotation of the plane along latitudes ($\beta$-effect). This instability is similar to the repelling regime of the zonal steady state in our model and the results of \cite{Holton2013Dynamic} are compared to ours in the following.

The first condition in \cite{Holton2013Dynamic} for instability to occur is
\begin{align}
    \overline{u}_{\rm T} > \frac{\beta}{\lambda_{\rm R}^2}.
\end{align}
This is also Phillips' criterion for baroclinic instability of a two-level model, see \cite{Phillips1954Energy}. Using the values in table \ref{tab:Fixed_parameters} we obtain a minimal shear of approximately $3.65\text{ m}\text{ s}^{-1}$, meaning that for a shear below this value the zonal steady state must always be attracting. In fact, it can be shown that for values of $\overline{u}_{\rm T}$ below this level our model is always in the attracting region of the zonal steady state and as the shear is increased this state becomes repelling.

According to \cite{Holton2013Dynamic}, the minimum value of $\overline{u}_{\rm T}$ for which $P^0$ becomes repelling occurs when $k^2=\sqrt2\lambda_{\rm R}^2/2$, where $k$ is the meridional wavenumber. Using the relation $k=j\pi/W$, this condition becomes
\begin{align}
    j=\frac{2^{-1/4}W\lambda_{\rm R}}{\pi}\approx 5.6,
\end{align}
where we used again the values in table \ref{tab:Fixed_parameters}. Analysing our model yields exactly that value of $j$ at which the mean zonal shear is minimal. The corresponding minimum is $\overline{u}_{\rm T}\approx 4.1\text{ m}\text{ s}^{-1}$ and is similar to the value given by \cite{Holton2013Dynamic}. This value of $j$ also yields the wavenumber of maximal instability of the purely zonal flow.

The short wavelength cut-off observed for large eddy diffusion $A$ in figure \ref{fig:j_H_u1_Avisc} on the left is given in \cite{Holton2013Dynamic} by the critical wavelength
\begin{align}
    \rm L_{\rm c}=\frac{2\pi}{\lambda^2_{\rm R}} \approx 3\times10^3\text{ km}.
\end{align}
For waves shorter than this value, instability can not occur and the zonal steady state remains attracting. In our model this value corresponds to $j>6.6$ and matches roughly the value for which the zonal steady state becomes attracting again if the eddy diffusion is very high and the heating rate is approximately the standard value given in table \ref{tab:Varying_parameters}.

\section{Q factor}\label{sec:Q_factor}
The quality factor Q is a dimensionless parameter that is used in mechanics and electronics to describe the oscillatory behaviour of a damped system. In general, the Q~factor is defined as the ratio between the natural undamped frequency of a dynamical system and its damping coefficient. Therefore, a high Q~factor indicates a larger number of oscillations before the system reaches the attracting steady state, in contrast to a low Q~factor indicating that the damping is more dominant and the attracting steady state is reached after fewer oscillations. A Q~factor below $1/2$ corresponds to an overdamped system that does not oscillate at all when displaced from its steady state, but returns to it by exponential decay.

This concept can be applied to the theory of damped oscillators, which satisfy the linear equation
\begin{align}
    \ddot x + \lambda \dot x + \omega^2_0 x = 0, \label{eq:damped_oscillator}
\end{align}
where $\omega_0^2$ is the frequency of free oscillations and $\lambda$ the damping coefficient. The general solution of equation (\ref{eq:damped_oscillator}) is
\begin{align}
    x=A\exp\left(-\frac{\lambda}{2}t\right)\exp\left(\pm i\left(\omega^2_0 -\frac{\lambda^2}{4}\right)^{1/2}t\right) = A\exp\left(pt\right), \label{eq:general_solution}
\end{align}
where A is a fixed complex amplitude and $p=-\lambda/2 \pm i\left(\omega^2_0-\lambda^2/4\right)^{1/2}$. The generally accepted definition of the Q~factor for the damped oscillator is
\begin{align}
    Q=\frac{\omega_0}{\lambda}. \label{eq:Q_factor_general}
\end{align}

We now consider a linearised system obeying the differential equation $\text{d} x/\text{d} t=\alpha x$.  This is associated with a linear damped oscillator if we take $p=\alpha=\alpha_{\rm r}+i\alpha_{\rm i}$, where $\alpha_{\rm r}<0$ is the real and $\alpha_{\rm i}$ the imaginary part of the eigenvalue. With this identification, we find $\alpha_{\rm r}=-\lambda/2$ and $\alpha_{\rm i}=\pm\left(\omega_0^2-\lambda^2/4\right)^{1/2}$ and hence the Q factor of a linearised system can be expressed as
\begin{align}
    Q = \frac{\left(\alpha_{\rm i}^2+\alpha_{\rm r}^2\right)^{1/2}}{-2\alpha_{\rm r}} = \frac{|\alpha|}{-2\alpha_{\rm r}}.\label{eq:Q_factor}
\end{align}
By construction, $Q\geq1/2$ for eigenvalues with a non-zero imaginary part, which is associated with the oscillatory behaviour of the system. For real eigenvalues, the Q~factor is exactly $1/2$ and the system is said to be critically damped. Similar to an over-damped system, the steady state is approached without oscillations.

\begin{figure}
\begin{center}
\subfigure{
\resizebox*{6.8cm}{!}
{\includegraphics{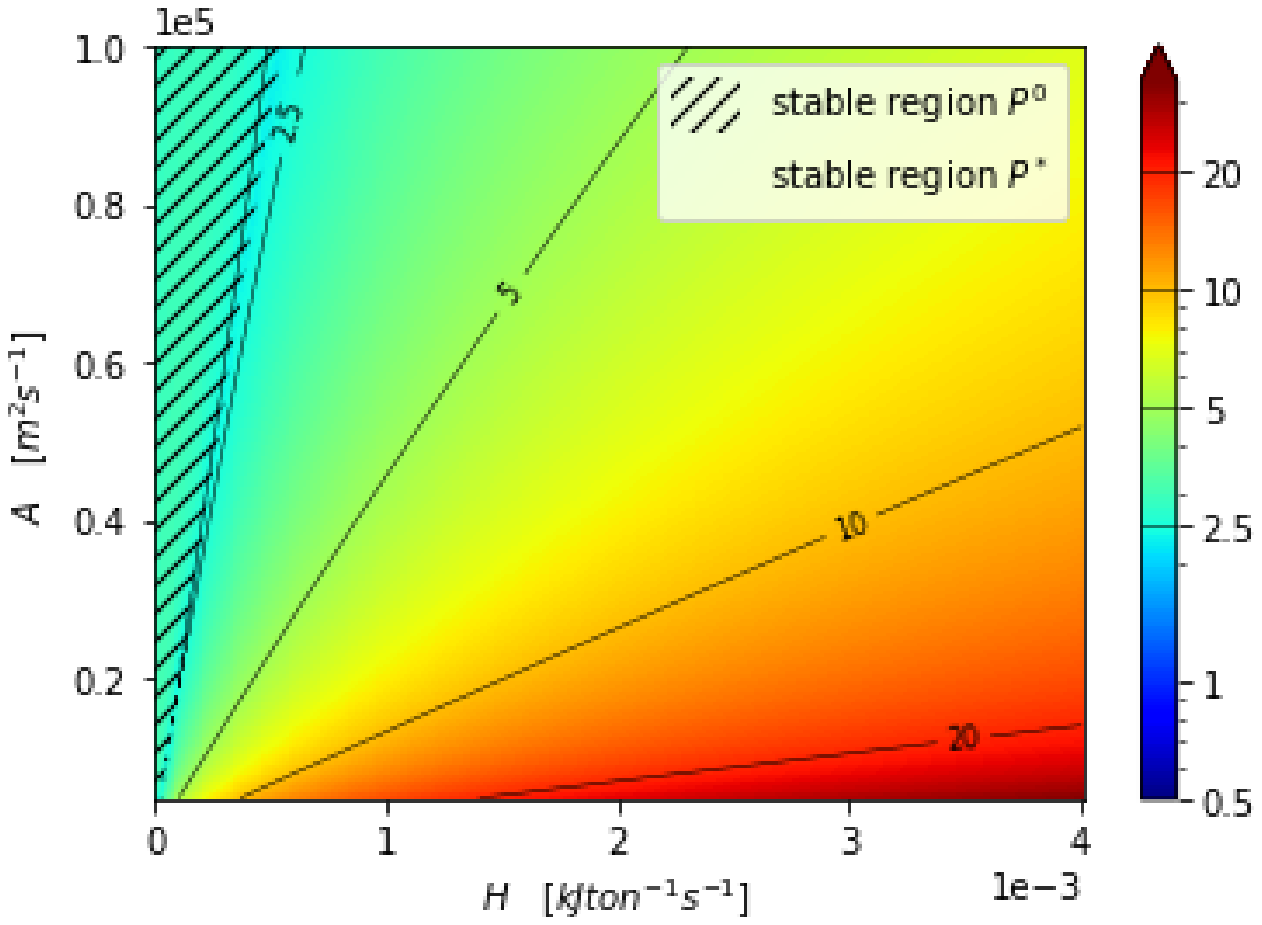}}}
\subfigure{
\resizebox*{7.1cm}{!}
{\includegraphics{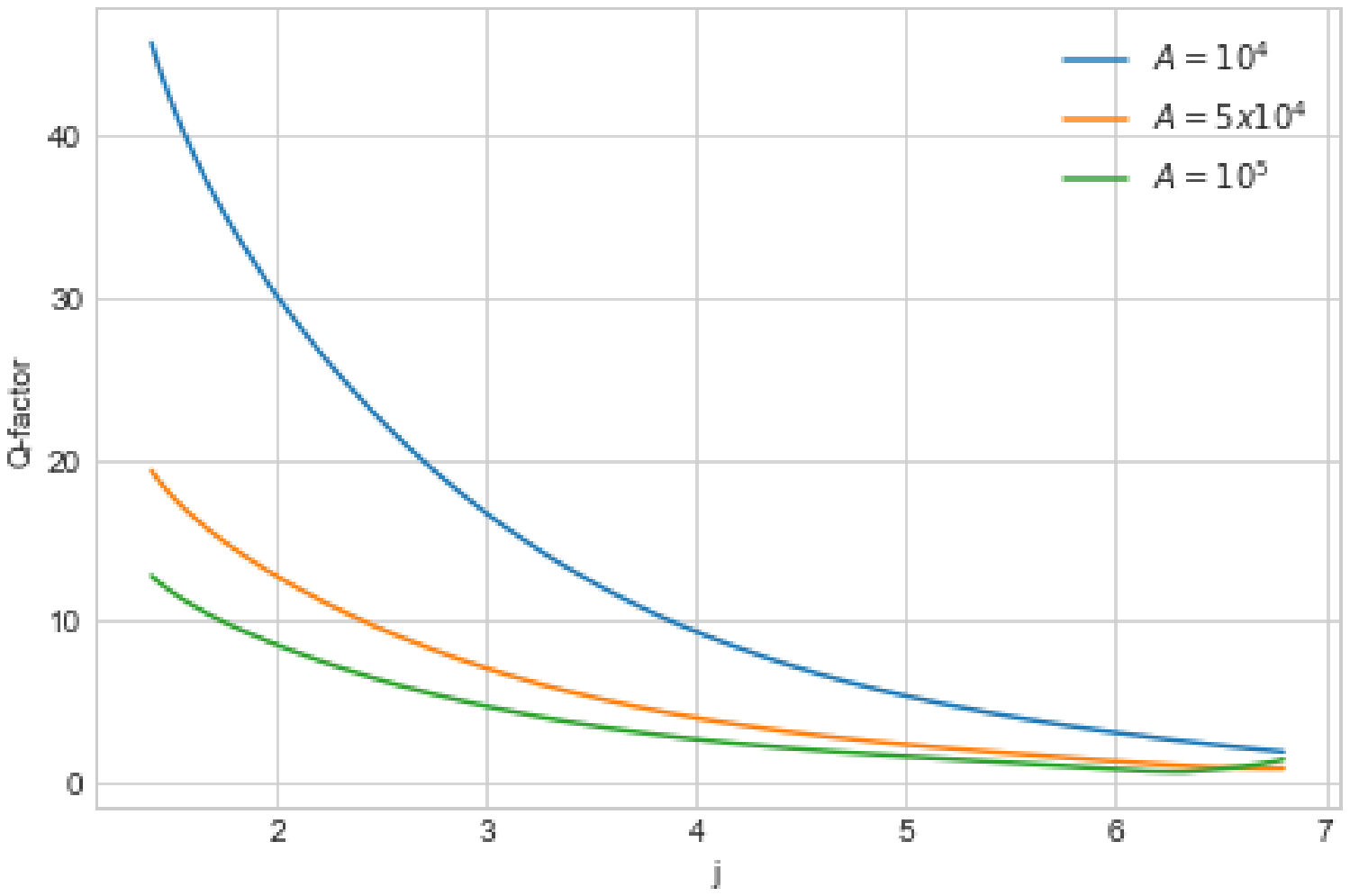}}}
\caption{\label{fig:Q_factor}
LEFT: Contours of the Q factor in the heating rate $H$ versus eddy diffusion $A$ plane, the dashed region is the parameter space where the zonal steady state $P^0$ is attracting, the non-dashed region where the eddy steady state $P^*$ is attracting, $\kappa=4\times10^{-6}\text{ s}^{-1}$,  $j=2.5$; RIGHT: Q factor as a function of $j$ for various values of the eddy diffusion $A$ in $\text{m}^{2}\text{ s}^{-1}$, $\kappa=4\times10^{-6}\text{ s}^{-1}$, $H=3.5\times10^{-3}\text{KJ ton}^{-1}\text{ s}^{-1}$.}
\end{center}
\end{figure}

The left hand side of figure \ref{fig:Q_factor} shows contours of the Q~factor in the heating rate $H$ versus eddy diffusion $A$ plane. As expected, Q is always greater than $1/2$ so the system oscillates everywhere in the considered parameter space. Furthermore, it can be seen that an increase of the eddy diffusion parameter $A$ leads to a decrease of the Q~factor. This corresponds to the fact that the diffusion acts as a damping and stabilizing factor of the system and leads to a reduction of the available energy. Therefore, the higher the diffusion, the less the initial transient relies on the eddy heat flux to compensate the heat imbalance. In contrast to that, a larger heating rate $H$ increases the Q~factor since more heat needs to be transported away from the equator and therefore more oscillatory cycles of the eddies are needed during the initial transient. The contours of Q in the heating rate versus surface friction $\kappa$ plane are similar (not shown) and can be interpreted in the same way.

On the right hand side of figure \ref{fig:Q_factor} the Q~factor is shown as a function of the eddy aspect ratio $j$ for different values of $A$. The highest Q~factor occurs for the smallest $j$, i.e. for the most symmetric eddy shape, and the larger $j$, the smaller is the Q factor. Hence, for eddies that are more elongated in the meridional direction, the system undergoes fewer oscillatory cycles before it reaches the attracting steady state. In physical terms, meridionally elongated eddies are more efficient in balancing the heat in the atmosphere than eddies close to symmetric. This also coincides with the findings in the earlier chapters. Apart from that, the dependence on the diffusion parameter $A$ is the same as in the contour plot on the left side.

\section{Discussion and conclusion}\label{sec:discussion}
We used Phillips' classical two-level quasi-geostrophic model on the $\beta$-plane \citep{Phillips1956Thegeneral} and performed a model reduction to develop a system of six ordinary differential equations describing the interaction between the mean flow and shear, and the eddy activity in the mid-latitude atmosphere. This has been achieved by imposing a specific shape on the eddy streamfunction and thus looking at the evolution of the amplitude of the atmospheric modes considered in our spectral projection.

Instead of taking the usual approach of performing a normal mode instability analysis, we used classical dynamical systems theory and determined two physically possible steady states, namely a zonal solution, where the flow is purely zonal and has no eddy contribution, and an eddy solution, where zonal flow and eddies coexist. The considered parameter space is divided into two separate regions where either the zonal or the eddy solution is attracting, while the alternative solution is unstable and repels nearby initial conditions.

The two competing states are in close correspondence with two possible physical  mechanisms that allow the atmosphere to reach a steady state in such a way that the imbalance in the diabatic heating between equator to the poles is compensated. Relatively weak imbalances in the diabatic heating can be balanced by diffusion and no eddies need to develop to flatten the meridional temperature gradient. Indeed, the flow is, in this case, baroclinically stable. In this state the mean flow increases for stronger forcing by diabatic heating, because the latter imposes a stronger temperature difference between low and high latitudes. This monotonicity is lost as soon as the heating reaches a certain threshold, where the eddies start growing to compensate the incoming heat because the damping effect of the eddy diffusion is no longer sufficient. The zonal flow is then baroclinically unstable so that eddies grow and reach a finite steady-state amplitude, in such a way that the resulting poleward eddy heat flux weakens the baroclinicity and therefore the mean flow. This process is referred to as baroclinic adjustment \citep{Stone1978Baroclinic}.

In this baroclinically unstable eddy steady state the zonal flow is independent of the external forcing by the heating. In contrast, the net poleward heat transport by the eddies and the eddy energy increase with incoming heat, so incoming energy in form of diabatic forcing of the zonal mean flow goes directly into the eddies in form of eddy kinetic energy, where it is dissipated. Therefore, the stability switch from zonal to eddy steady state describes a transfer of the location of dissipation of incoming energy. In terms of the Lorenz energy cycle, for sufficiently large heating, the zonal flow can no longer act as the reservoir for the incoming energy, so a new reservoir needs to develop, namely the eddies. Thus, the reduced model is exhibiting complete eddy saturation.

In contrast to \cite{Munday2013Eddy}, stating that eddy-resolving models of the ACC lead to a much lower sensitivity of the volume transport to increased forcing, \cite{Marshall2017Eddy} suggested the possibility to capture the physics of eddy saturation in models with parameterised eddies. Thereafter, \cite{Mak2017Emergent} used a highly idealised model configuration of the ACC with parameterised eddies to obtain a near total eddy saturation. For the atmosphere, \cite{Novak2018Baroclinic} found a model where the mean baroclinicity is nearly independent of the thermal forcing, whereas the eddy intensity responds more strongly to the forcing and is independent of the eddy dissipation. In this context, the model presented in this work is the first to exhibit complete eddy saturation in a model of the mid-latitude atmosphere with parameterised eddies.

Another aspect of the eddy saturation mechanism described in \cite{Marshall2017Eddy} is the increase of the volume transport with the bottom drag, and one would expect a similar effect of the surface friction in our model, dissipating energy from the system and therefore compensating for the incoming heat. This would lead to a stability switch at higher values of the heating and to an increase of the zonal wind speed. However, this effect can only be seen for unrealistically high values of the surface friction, perhaps indicating a limitation of our model. Due to the two-level setup the model is able to develop a high wind shear induced by the incoming heating by increasing the upper level zonal wind but maintaining a very low surface wind. Hence, the surface friction can not directly compensate for the incoming heating in the model, although in the atmosphere the heating actually induces stronger surface winds.

In addition to the zonal wind being independent of the incoming heating and the surface friction (for realistic values) in the baroclinically unstable eddy steady state, the wind speed is independent of the eddy diffusion as well. Instead, an increased diffusion leads to a decrease of eddy energy, indicating the damping effect of eddy dissipation which is known as \textit{dissipative control} (e.g. \cite{Novak2018Baroclinic}), and is in accordance with \cite{Marshall2017Eddy}, stating that the equilibrium volume transport is controlled by the ACC requiring sufficiently strong vertical shear induced by external forcing to overcome the stabilising role of the eddy dissipation.

Despite the simplicity of the model and the limiting assumptions made on the shape of the eddies, our model exhibits complete eddy saturation, a regime of the mid-latitude atmosphere that appears to be supported by numerical models. The eddy saturation mechanism is explained as a switch of stability from a zonal mean state to a mean state with finite eddy amplitude. This work is of relevance to climate modeling, where the response of the mid-latitude storm tracks to changes in the diabatic heating is still not fully understood. 

\section*{Acknowledgement}
Melanie Kobras is supported by the
U.K. Engineering and Physical Sciences Research Council (Grant EP/R513301/1). Valerio Lucarini acknowledges the support received from the EPSRC project EP/T018178/1 and from the EU Horizon 2020 project TiPES (grant no. 820970).

\bibliographystyle{tfcad}
\bibliography{References}

\appendix

\section{Additional evolution equations}\label{sec:Additional_equations}

The evolution equations for the mean meridional kinetic energy $\overline{v'^2_{\rm m}}$, the temperature variance $\overline{v'^2_{\rm T}}$ and the cross-correlation between temperature and geopotential $\overline{v'_{\rm m}v'_{\rm T}}$ are
\begin{subequations}
\begin{align}
        \frac{\upartial\overline{v'^2_{\rm m}}}{\upartial t} &=  \frac{2k^2(\lambda^2_{\nabla}-\lambda^2_y)}{\lambda^2_{\nabla}}\overline{u}_{\rm T}\overline{v'_{\rm m}\psi'_{\rm T}} - (2A\lambda^2_{\nabla}+\kappa)\overline{v'^2_{\rm m}} + 2\kappa\overline{v'_{\rm m}v'_{\rm T}}, \label{eq:v^2_m}\\
        \frac{\upartial\overline{v'^2_{\rm T}}}{\upartial t} &= \frac{2k^2(\lambda^2_y+\lambda^2_{\rm R}-\lambda^2_{\nabla})}{\lambda^2_{\nabla}+\lambda^2_{\rm R}}\overline{u}_{\rm T}\overline{v'_{\rm m}\psi'_{\rm T}} - \left(2A\lambda^2_{\nabla}+\frac{2\kappa \lambda^2_{\nabla}}{\lambda^2_{\nabla}+\lambda^2_{\rm R}}\right)\overline{v'^2_{\rm T}} + \frac{\kappa \lambda^2_{\nabla}}{\lambda^2_{\nabla}+\lambda^2_{\rm R}}\overline{v'_{\rm m}v'_{\rm T}},\label{eq:v^2_T}\\
        \frac{\upartial}{\upartial t}\overline{v'_{\rm m}v'_{\rm T}} &= \frac{\lambda^2_{\rm R}k^2}{\lambda^2_{\nabla}(\lambda^2_{\nabla}+\lambda^2_{\rm R})}\left(\beta+\lambda^2_y\overline{u}_{\rm m}\right)\overline{v'_{\rm m}\psi'_{\rm T}} - \left(2A\lambda^2_{\nabla}+\frac{\kappa(3\lambda^2_{\nabla}+\lambda^2_{\rm R})}{2(\lambda^2_{\nabla}+\lambda^2_{\rm R})}\right)\overline{v'_{\rm m}v'_{\rm T}} + \kappa\overline{v'^2_{\rm T}} \nonumber \\
        &\qquad + \frac{\kappa \lambda^2_{\nabla}}{2(\lambda^2_{\nabla}+\lambda^2_{\rm R})}\overline{v'^2_{\rm m}} \label{eq:v_m_v_T}.
\end{align}
\end{subequations}

\section{Dimensionless constants of model (\ref{eq:dMdt})-(\ref{eq:dXdt})}\label{sec:dim-less constants}
\begin{align}
    \alpha=\frac{\lambda_{\rm R}}{\beta}\kappa&, \quad
    \gamma=\frac{\lambda^2_y}{\lambda^2_y+\lambda^2_{\rm R}}, \quad
    \delta= \frac{\lambda^2_{\rm R}}{\hat{j}(\hat{j}\lambda^2_y+\lambda^2_{\rm R})}, \quad
    \epsilon= \frac{3\hat{j}\lambda^2_y+\lambda^2_{\rm R}}{\hat{j}\lambda^2_y+\lambda^2_{\rm R}}, \quad
    \nu=\frac{\lambda_{\rm R}\lambda^2_y A}{\beta}, \nonumber\\
    \zeta&=\frac{\hat{j}\lambda^2_y}{\hat{j}\lambda^2_y+\lambda^2_{\rm R}}, \quad
    \mu=\frac{\lambda^2_{\rm R}+(1-\hat{j})\lambda^2_y}{\hat{j}\lambda^2_y+\lambda^2_{\rm R}} \quad
    \eta= \frac{\lambda^5_{\rm R} R}{\beta^2 f_0 c_{\rm p} W(\lambda^2_y+\lambda^2_{\rm R})},
    \label{eq:QG13}
\end{align}
where
\begin{align}
    \hat{j}=\frac{j^2+1}{2}, \
    a=\nu+\frac{\alpha}{2}, \
    b=\nu+\gamma\alpha, \
    c=2\hat{j}\nu+\frac{\alpha\epsilon}{2}, \
    d=2\hat{j}\nu+\alpha, \
    e=2\hat{j}\nu+2\alpha\zeta.
    \label{eq:QG14}
\end{align}

\section{Parameter space}\label{sec:Parameter_space}
\begin{table}[h]
\tbl{Fixed parameter values for stability analysis.}
{\begin{tabular}{lccccccc}\toprule
Parameter
& $\lambda^2_{\rm R}$ & $\lambda^2_y$ & $\beta$ & $R$ & $f_0$ & $c_{\rm p}$ & $W$ \\
\midrule
Value & $4.39\times10^{-12}$ & $1.97\times10^{-13}$ & $1.6\times10^{-11}$ & $287$ & $10^{-4}$ & $1004$ & $10^{7}$ \\
Unit & $\text{m}^{-2}$ & $\text{m}^{-2}$ & $\text{m}^{-1}\text{ s}^{-1}$ & $\text{J K}^{-1}\text{ kg}^{-1}$ & $\text{s}^{-1}$ & $\text{J K}^{-1}\text{ kg}^{-1}$ & $\text{m}$\\
\bottomrule
\end{tabular}}
\label{tab:Fixed_parameters}
\end{table}

\begin{table}[h]
\tbl{Varying parameter values for stability analysis.}
{\begin{tabular}{lcccc}\toprule
Parameter
& $\kappa$ & $A$ & $H$ & $j$ \\
\midrule
Minimum & $10^{-7}$ & $1$ & $0$ & $0.1$ \\
Standard & $4\times10^{-6}$ & $10^{5}$ & $3.5\times10^{-3}$ & - \\
Maximum & $8\times10^{-6}$ & $10^{5}$ & $4\times10^{-3}$ & $6.8$ \\
Unit & $\text{s}^{-1}$ & $\text{m}^{2}\text{ s}^{-1}$ & $\text{KJ ton}^{-1}\text{ s}^{-1}$ & - \\
\bottomrule
\end{tabular}}
\label{tab:Varying_parameters}
\end{table}

\section{Other variables of eddy steady state}\label{sec:eddy_steady_state}
Non-dimensional eddy steady state variables of the mean meridional kinetic energy $K^*$, the temperature variance $V^*$ and the cross-correlation between temperature and geopotential $X^*$:
\begin{align}
    K^* &= \frac{2^2j^2(j^2-1)\lambda^2_y}{(j^2+1)d\lambda^2_{\rm R}}S^*T^* + \frac{2\alpha}{d}X^*,&
    V^* = \frac{2^2j^2\mu\lambda^2_y}{e\lambda^2_{\rm R}}S^*T^* + \frac{\alpha\zeta}{e}X^*,\nonumber \\
    X^* &= \left(\frac{2j^2\alpha\lambda^2_y g}{a\lambda^2_{\rm R}}S^* + j^2\delta\right)\left(c-\alpha\zeta\frac{d+e}{de}\right)^{-1}T^*,& \label{eq:eddy_state_2}
\end{align}
with constants
\begin{align*}
    B &= 2\alpha\delta(g-f)\left(c-\alpha^2\zeta\frac{d+e}{de}\right)^{-1},\quad
    C = \frac{\delta^2\lambda^2_{\rm R}}{2\lambda^2_y}\left(c-\alpha^2\zeta\frac{d+e}{de}\right)^{-1} +c,\\
    D &= 8\frac{\lambda^2_y}{\lambda^2_{\rm R}}\left(2a^2\mu\left(1-\frac{1}{\hat{j}}\right)\frac{d+e}{de} + \alpha^2fg\left(c-\alpha^2\zeta\frac{d+e}{de}\right)^{-1}\right),\\
    f &= \frac{2a\mu}{d} + (1-\frac{1}{\hat{j}})\frac{a\zeta}{e} - \delta,\quad
    g = \delta + \frac{(j^2-1)a\zeta}{(j^2+1)d} + \frac{2a\mu}{e}.
\end{align*}
Here, $B$, $C$ and $g$ are positive for the whole parameter space given by tables \ref{tab:Fixed_parameters} and \ref{tab:Varying_parameters}, $D$ and $f$ are non-negative for $j$ larger than 1 and for the positivity of $D$ additionally $\kappa \geq 3\times10^{-6}\text{ s}^{-1}$ is required.

\end{document}